\documentclass[%
reprint,
superscriptaddress,
 amsmath,amssymb,
 aps,
]{revtex4-2}

\usepackage{graphicx}
\usepackage{dcolumn}
\usepackage{bm}
\usepackage{siunitx}

\begin{document}

\preprint{APS/123-QED}

\title{A mechanism for electrostatically generated magnetoresistance in chiral systems without spin-dependent transport}

\author{Sytze H. Tirion}
\email{s.h.tirion@rug.nl}
\affiliation{Zernike Institute for Advanced Materials, University of Groningen, NL-9747AG Groningen, The Netherlands}

\author{Bart J. van Wees}
\affiliation{Zernike Institute for Advanced Materials, University of Groningen, NL-9747AG Groningen, The Netherlands}

\date{\today}

\begin{abstract}
Significant attention has been drawn to electronic transport in chiral materials coupled to ferromagnets in the chirality induced spin selectivity (CISS) effect. A large magnetoresistance (MR) is usually observed which is widely interpreted to originate from spin (dependent) transport. However, there are severe discrepancies between the experimental results and theoretical interpretations, most notably the apparent failure of the Onsager reciprocity relation in the linear response regime. We provide an alternative explanation for the mechanism of the two terminal MR in chiral systems coupled to a ferromagnet. For this we point out that it was observed that the electrostatic contact potential of chiral materials on a ferromagnet depends on the magnetization direction and chirality. In our explanation this causes the transport barrier to be modified by the magnetization direction, already in equilibrium, in the absence of a bias current. This strongly alters the charge transport through/over the barrier, not requiring spin transport. This provides a mechanism that allows the linear response resistance to be sensitive to the magnetization direction and also explains the failure of the Onsager reciprocity relations. We propose experimental configurations to confirm our alternative mechanism for MR.

\end{abstract}

\maketitle

\section{\label{sec:level1}Introduction}

In recent years the chirality of materials has been connected to the electron spin in an effect titled the chirality-induced spin selectivity (CISS) effect \cite{Evers2022,Aiello2022,Yang2021a,yan2023}. It is a collective term for the interpretation of a diverse range of experiments where chirality is assumed to give rise to various spin-dependent effects. A major direction in the field of CISS is formed by experiments on electronic transport through chiral systems coupled to a ferromagnet, which usually show a large magnetoresistance (MR) \cite{Mishra2020,Mondal2021,Malatong2023}. Although a wide range of theories have been proposed, \cite{Yang2019,Yang2020a,Yang2021,Dalum2019,Fransson2021,Alwan2021,Hedegard2023,Klein2023,xiao2023nonreciprocal}, the discrepancies between the experimental results and the theories remain large \cite{Evers2022}. It is crucial to understand the origin of the MR that is experimentally observed for chiral systems connected to a ferromagnet, both for understanding of fundamental physics, as well as possible applications of chiral systems in the field of spintronics.

Currently, the large MR is assumed to originate from spin-dependent transport in, or into the chiral system. The chiral system is presumed to be a spin polarizer/filter which, together with the ferromagnet as spin polarizer/analyzer, will give a modification of the spin transport when either the chirality of the system is changed or the magnetization direction is reversed. This then causes an MR similar to giant magnetoresistance (GMR) \cite{Bass2016} or tunnel magnetoresistance (TMR) \cite{Lee2017} in purely ferromagnetic systems.

Here, we first review the experimental results of the MR measured for chiral systems connected to a ferromagnet and identify the incompatibility of them in terms of spin transport. Then, we provide an alternative explanation for the origin of the MR of chiral systems coupled to a ferromagnet. We argue that experiments measured the electrostatic contact potential of the system in equilibrium that is altered by either reversal of the magnetization direction or chirality \cite{Theiler2023,Ghosh2020,Abendroth2019}. We investigate the implications of this for the modification of the potential profiles of a Schottky barrier-like junction where the chiral system is in direct contact with a ferromagnetic metal, as well as the case where they are separated by a tunnel barrier. We find that the barrier height and the energy bands within the screening length are adjusted and thus can modify the charge transport. This gives an MR which does not dependent on spin dependent transport.

Although we do not understand its origin, the fact that the electrostatic equilibrium properties of a chiral system coupled to a ferromagnet are modified by the magnetization direction also causes the Onsager reciprocity relations to fail and gives an MR in the linear response.

A recent theoretical discussion by \citeauthor{xiao2023nonreciprocal} \cite{xiao2023nonreciprocal} of the large MR also discuss a modification of the charge transport barrier between the chiral system and the ferromagnet \cite{Adhikari2022,xiao2023nonreciprocal,yan2023}. However, it is argued that this is due to bias current induced charge trapping that can only result in an MR in the non-linear bias regime \cite{Yang2020a,Huisman2021}. It therefore cannot explain the experimental observation of MR in linear response.

\section{CISS magnetoresistance, overview of experimental techniques and results}
\label{sec:tech_and_results}

The electronic transport in chiral systems coupled to ferromagnets is studied by two main experimental techniques.

Local scanning probe techniques (conductive atomic force microscopy \cite{Lu2019,Mishra2020b,Mishra2020,Kulkarni2020,Mondal2021,Zhu2021,Amsallem2023,Labella2023,Metzger2023,Malatong2023,Bian2023,Zhang2023} and scanning tunneling microscopy \cite{Nguyen2022,Safari2023}) where a thin film of chiral material (in most cases molecules) is absorbed on a conducting substrate and a local probe is used to measure the two terminal voltage-current (VI) characteristics of the chiral system with typical currents of nanoamperes, for applied biases up to several Volts. Either the substrate or the probe is ferromagnetic and the VI characteristics of the chiral system are measured with the magnetization pointing either up or down. The (averaged) VI curves of the up and down magnetization are compared, and the MR is calculated as a function of the applied bias. Comparatively large MR values, typically above 50\% and in some cases close to 100\% \cite{Kulkarni2020,Bian2023}, are reported. Importantly, in many reports the measured MR is (almost) bias independent \cite{Torres-Cavanillas2020,Kulkarni2020,AlBustami2022a,Bian2023,Labella2023,Malatong2023} for a bias up to several Volts.

Alternatively, junction geometries are used \cite{Lu2019,Suda2019,Mishra2020,Mishra2020b,Torres-Cavanillas2020,Liu2020,AlBustami2022a,Das2022,Adhikari2022,Hossain2023} where the chiral layer is sandwiched between two microfabricated electrodes, one ferromagnetic with an oxide barrier in between it and the chiral system. An external magnetic field orients the magnetization. In these two terminal experiments, either the VI characteristic is measured, and the bias dependence of the MR is calculated. In most cases the reported MR from the junctions is an order of magnitude smaller compared to the local scanning probe techniques, but recently some junctions also show  high MR values \cite{Bian2023,AlBustami2022a}.

Compared to the two terminal geometry, we note that it is difficult to perform four terminal measurements on the chiral systems that have been studied so far. However, the electronic transport of chiral solid state crystal system was studied by four terminal measurements \cite{Inui2020,Shishido2021,Shiota2021}, see Appendix \ref{sec:App_Linear_Resp_CISS}. For recent theoretical overviews of spin selectivity of chiral solid state system see Ref \cite{Sawinska2023}\& \cite{yang2023chirality}.

\section{Problems with the spin transport interpretation of two terminal magnetoresistance}

\begin{figure*}
\centering
\includegraphics[width=0.95\textwidth]{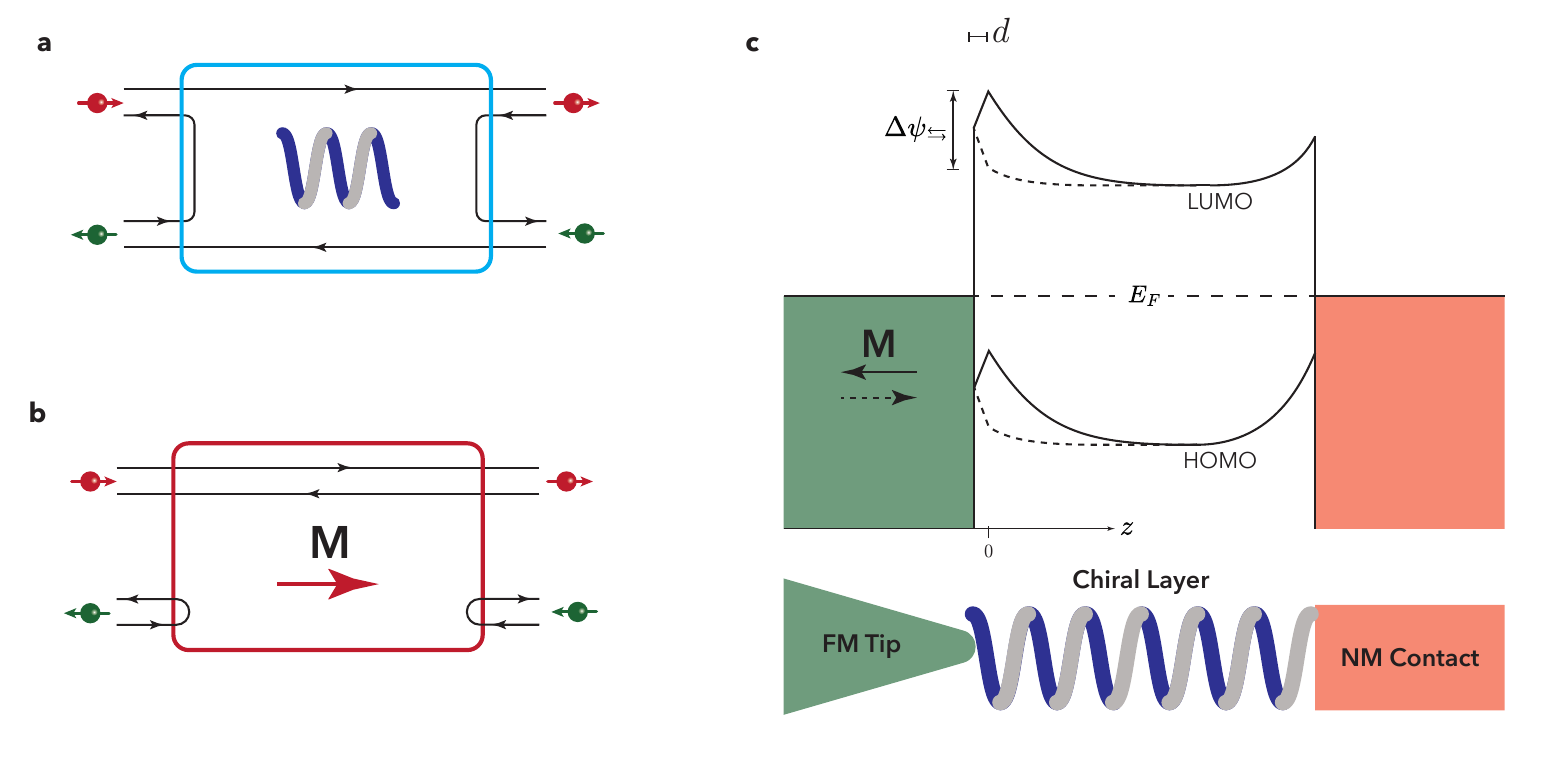}
\caption{\label{fig:workfunc_modulation} (a) Elementary spin-dependent transmission and reflection model for a chiral system. The ``favoured'' electron spin is transmitted and the ``unfavoured'' electron spin is reflected but has to spin-flip \cite{Yang2019,Yang2020a} because directional spin transmission always requires a spin-flip reflection process to avoid spin currents in equilibrium. (b) A ferromagnet gives rise to spin-dependent transport but this is fundamentally different because the preferred spin is set by the magnetization direction, not the propagation direction. (c) An energy level diagram of a chiral system contacted by a ferromagnet tip (FM Tip) and a metal substrate (NM) to illustrate the alternative origin for the MR. At the interface between the tip and the chiral system a Schottky like barrier is formed. The magnetization direction affects the contact potential $\Delta \psi_{\leftrightarrows}$ that is build up over a distance $d$. This then modifies the electrostatic potential profile of the chiral molecules.}
\end{figure*}

The interpretation of the experimental results assumes that electron transport in chiral systems is spin-dependent, where the chirality sets the preferred spin direction which is coupled to a given propagation direction. However, the measured MR is often directly interpreted in terms of the spin polarization of the transmitted electrons. It is crucial to distinguish the MR \cite{Yang2020a,Wolf2022,Liu2023} generated by the entire circuit (chiral system + ferromagnet) from mechanisms that generate the spin-dependent electron transport in the chiral systems themselves \cite{Matityahu2016,Utsumi2020}. In this work we assume the presence of directional spin-dependent transmission and reflection in the chiral systems itself. However, despite this assumption, below we identify five main problems with the common interpretation of the measured MR in terms of spin-dependent transport.

\subsection{Unknown mechanism for spin injection and detection by the ferromagnet}
In past decades, significant effort was dedicated to electrical injection/detection of spin-polarized electrons \cite{Schmidt2000, Schmidt2005} into metals and semiconductors, which involves two different transport mechanisms. The first is spin-dependent conductivity determined by the polarization of the ferromagnet at the Fermi energy, which is the mechanism of giant magnetoresistance (GMR) \cite{Bass2016}. The second is spin-dependent tunneling proportional to the density of states at the Fermi energy commonly used in tunnel magnetoresistance (TMR) \cite{Lee2017}. However, for most chiral systems it is unclear which of these mechanisms could be responsible for the assumed spin injection/detection from the ferromagnet.

\subsection{Conductivity mismatch}
The (low bias) injection of spin-polarized electrons by a ferromagnetic metal into semiconductors or other low-conductivity materials is strongly hindered by the large difference in conductivity, know as the ``conductivity mismatch'' \cite{Schmidt2000}. This problem can be overcome by impedance matching the ferromagnet and the semiconductor e.g. by a tunnel junction. This will make spin-dependent tunneling a possible mechanism for spin injection and detection from a ferromagnet into a chiral system. However, it is unaddressed how the conductivity mismatch between the ferromagnetic metal and the chiral system is overcome. This is especially relevant since most of the chiral systems studied are molecules with comparatively low conductivity, making the mismatch problem only more significant.

\subsection{Large magnetoresistance}
The reported MR for chiral systems coupled to a ferromagnet is typically above 50\% and approaches 100\% in a number of recent experimental results \cite{Kulkarni2020,Bian2023}. A conventional TMR spin valve consist of two ferromagnetic electrodes and operates based on spin-dependent tunneling for which the Julière model connects the spin polarization of the density of states of the ferromagnets, $P_{FM}$, to the MR \cite{Liu2023}. Using this model, it can be shown that,

\begin{equation}
	MR = \frac{I_{\uparrow} - I_{\downarrow}}{I_{\uparrow} + I_{\downarrow}} = P_{FM} P_{Chiral},
	\label{eq:Jull}
\end{equation}
where $I_{\uparrow \downarrow}$ is the measured current with magnetization direction indicated by the subscript and $P_{Chiral}$ the``spin polarization'' of the chiral system. This clearly shows that the total MR can never exceed the polarization of the ferromagnet, even when we assume a perfect spin polarization of the chiral system ($P_{Chiral} = 1$), as was recently also pointed out by \citeauthor{Liu2023} \cite{Liu2023} and others. Note that here the MR definition commonly used for CISS is applied, which is different from the conventional definition of the TMR ratio applied in the Julière model.

\subsection{Bias voltage dependence of the magnetoresistance}
Theoretical effort is dedicated to explain the high MR in terms of the spin polarization of transmitted electrons from the chiral system, arguing why $P_{Chiral}$ could be close to 1. However, as far as we know the theoretically predicted spin polarization of both the chiral systems as well as for ferromagnets, is strongly energy-dependent and hence bias dependent \cite{Matityahu2016,Liu2021,Dalum2019,Utsumi2020,Fransson2021,Alwan2021,Klein2023,VanRuitenbeek2023,Huisman2021,Huisman2022,Huisman2023}. This is in stark contrast with the experimental results which shows in many reports, an (almost) bias independent MR \cite{Torres-Cavanillas2020,Kulkarni2020,AlBustami2022a,Bian2023,Labella2023,Malatong2023}, for a bias range of several Volts. This also makes the experimentally observed MR an 'even' function of the bias, meaning it does not change sign when the bias is reversed. This does not agree with expected symmetry which should be 'odd' in bias, meaning a change in sign of the MR when bias is reversed \cite{Yang2020a,Huisman2021}, see Appendix \ref{sec:MR_CISS_non-linear}.

\subsection{Failure of the Onsager reciprocity relations}
In linear response the charge and spin transport parameters can be fully determined by the equilibrium properties of the sample, making it adhere to the Onsager reciprocity relations (see Appendix \ref{sec:App_Linear_Resp}). 

It is important to disentangle the MR generated by chiral systems coupled to a ferromagnet from the possible spin-dependent transport in the chiral system itself \cite{Waldeck2021}, because these are different concepts. FIG.~\ref{fig:workfunc_modulation}(a)\&(b) shows the spin-dependent transmission model developed by \citeauthor{Yang2019} \cite{Yang2019,Yang2020a,Yang2021}, that includes both the spin-dependent transmission and the reflection processes of chiral and ferromagnetic systems. It was found that the linear response resistance is unaffected by reversal of the magnetization, see FIG.~\ref{fig:App} in Appendix \ref{sec:App_Linear_Resp_CISS}. Only in the nonlinear regime a MR can be detected which requires a combination of energy dependent transport and energy relaxation (e.g. possible electron-electron and electron-phonon interactions \cite{Fransson2019,Fransson2021,Huisman2022,Huisman2023}), Appendix \ref{sec:MR_CISS_non-linear}. 

However, several experimental results for a wide range of chiral systems \cite{Torres-Cavanillas2020,Clever2022,AlBustami2022a,Bian2023,Malatong2023,Metzger2023} now show a clear difference in the linear response resistance when the magnetization is reversed, which is at odds with Onsager reciprocity. This makes the origin of the MR reported for chiral systems coupled to a ferromagnet one of the main enigmas in the field of CISS.

Based on the discussion above we argue that the origin of the MR reported in (two terminal) CISS transport experiments is not due to a (modification of the) spin injection/transport. Instead, we provide an alternative explanation for the origin of the MR in terms of a chirality dependent modification of the contact potential by the magnetization, not involving spin transport.

\section{Magnetochiral contact potential modification}

\begin{figure}[h!]
\centering
\includegraphics[width=0.48\textwidth]{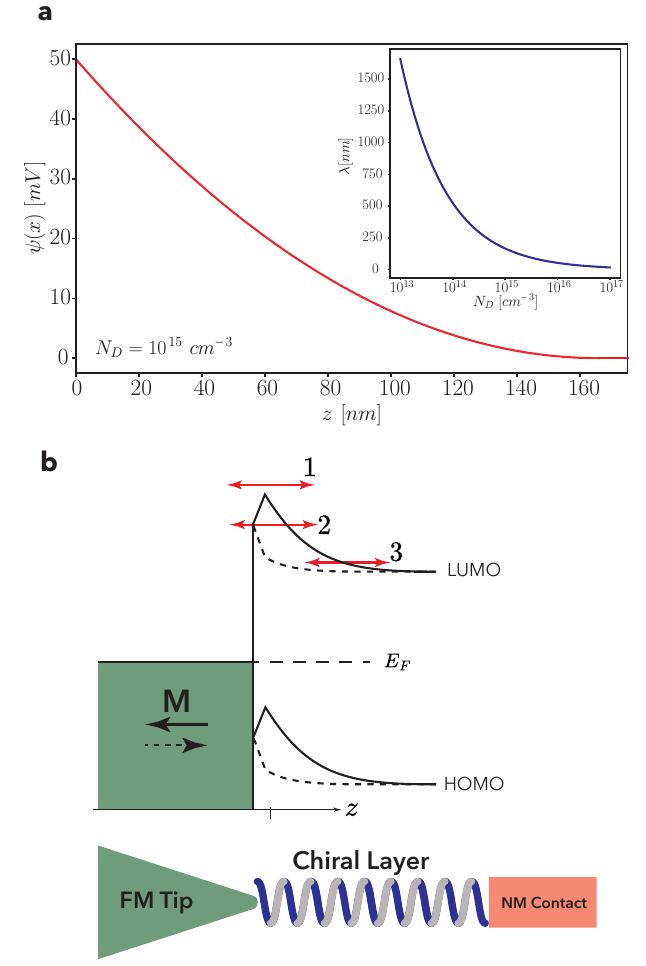}
\caption{\label{fig:workfunc_modulation_graph} (a) The calculated potential $\psi(z)$ for a contact potential change of \SI{50}{\milli \volt}. We use $N_D =$ \SI{e15}{\per\centi\meter\cubed} and $\epsilon_r = 5$. The inserted graph shows the screening length $\lambda$ as function of the doping concentration on a semi-logarithmic x-axis. (b) Energy level diagram of a chiral system in contact with a ferromagnet. When a bias is applied the (modified) transport can occur via three mechanisms: 1. Thermionic emission 2. Tunneling through the barrier 3. Transport within the screening length.}
\end{figure}

Recent experimental results report a modification of the (electrostatic) equilibrium properties of a chiral system coupled to a ferromagnet in the absence of a bias current \cite{Abendroth2019,Ghosh2020,Theiler2023}. It was found that the contact potential of chiral molecules on a ferromagnet is modified by reversal of either the magnetization direction or by interchanging the chirality of the system. \citeauthor{Theiler2023} \cite{Theiler2023} recently reported a shift of the contact potential measured by Kelvin probe force microscopy (KPFM) of \SI{50}{\milli \volt} when the direction of the magnetization is reversed, which is similar in magnitude previously reported by KPFM measurements \cite{Ghosh2020} and photo-emission experiments \cite{Abendroth2019}. We point out that no electron transport is involved in these measurements of the contact potential shifts. The large experimental timescales, in our opinion, exclude transient effects, justifying that this is an equilibrium phenomenon. The magnitude of the contact potential shift of $\sim$ \SI{50}{\milli \volt}, is large compared to $k_bT$, where $k_b$ is the Boltzmann constant and $T$ is the (room) temperature, which makes the change in contact potential relevant for the (low bias) electron transport and the MR.

Another recent experiment by \citeauthor{Volpi2023} \cite{Volpi2023} has investigated field effect transistors with a chiral semiconductor channel and ferromagnetic source and drain electrodes. It was found that the threshold gate voltage is strongly modified by the magnetization direction, highlighting the significance of electrostatic effects for chiral systems coupled to a ferromagnet.

Taking the modification of the contact potential by the magnetization direction as a starting point, we give an alternative explanation for the origin of the MR reported for chiral systems coupled to a ferromagnet, for which we use a semiconductor analogy. We assume a chirality dependent modification of the contact potential when the magnetization direction is reversed, which can be seen as a change of the electron affinity of the molecules at the interface with the ferromagnet. This is strongly influences the electrostatic potential landscape of the junction between a ferromagnet a the chiral molecular system, for which we use a semiconductor band picture with the HOMO and LUMO bands. Therefore, the electronic charge transport (but not the spin transport) through the ferromagnet/molecular junction is affected by the magnetization direction.

For a direct contact between metallic ferromagnet and the chiral molecular system a Schottky-type barrier, where the HOMO/LUMO bands bend, is expected. In FIG.~\ref{fig:workfunc_modulation}(c), the effect of the contact potential modification on the barrier between the ferromagnet and a molecular system is depicted. In equilibrium, with a constant Fermi energy, the magnetization direction changes the contact potential and, therefore modifies the electrostatic potential and the barrier height by $\Delta\psi_\leftrightarrows $, over a distance given by the screening length $\lambda$.

To check the feasibility of this explanation, we calculate the areal electron density, $N$, that is required to build up a potential difference of $\Delta\psi_{\leftrightarrows} = $ \SI{50}{\milli\volt} \cite{Theiler2023} over a typical length scale of a molecule $d = $ \SI{1}{\nano \meter}, using a parallel plate capacitor approximation,

\begin{equation}
	N = \frac{\epsilon_s \Delta \psi_{\leftrightarrows}}{e d},
\end{equation}

where $\epsilon_s = \epsilon_r \epsilon_0$ is the permittivity  and $e$ the electron charge. With $\epsilon_r = $ \SI{5}{}, we find that $N = $ \SI{e-2}{\per \nano \meter \squared}, which corresponds to \SI{2.5e-3}{} electron charge per molecule for a molecular packing density of \SI{4}{\per \nano \meter \squared}. This indicates that only a small fraction of charge shift per single molecule is required to build up a potential of \SI{50}{\milli\volt}.

In a Schottky like junction the HOMO and LUMO bands bend when they return to their bulk position. This happens over the length scale, $\lambda$, which depends on the shift of contact potential. We estimate the length scale $\lambda$ with the the Poisson-Boltzmann equation,

\begin{figure*}
\centering
\includegraphics[width=0.99\textwidth]{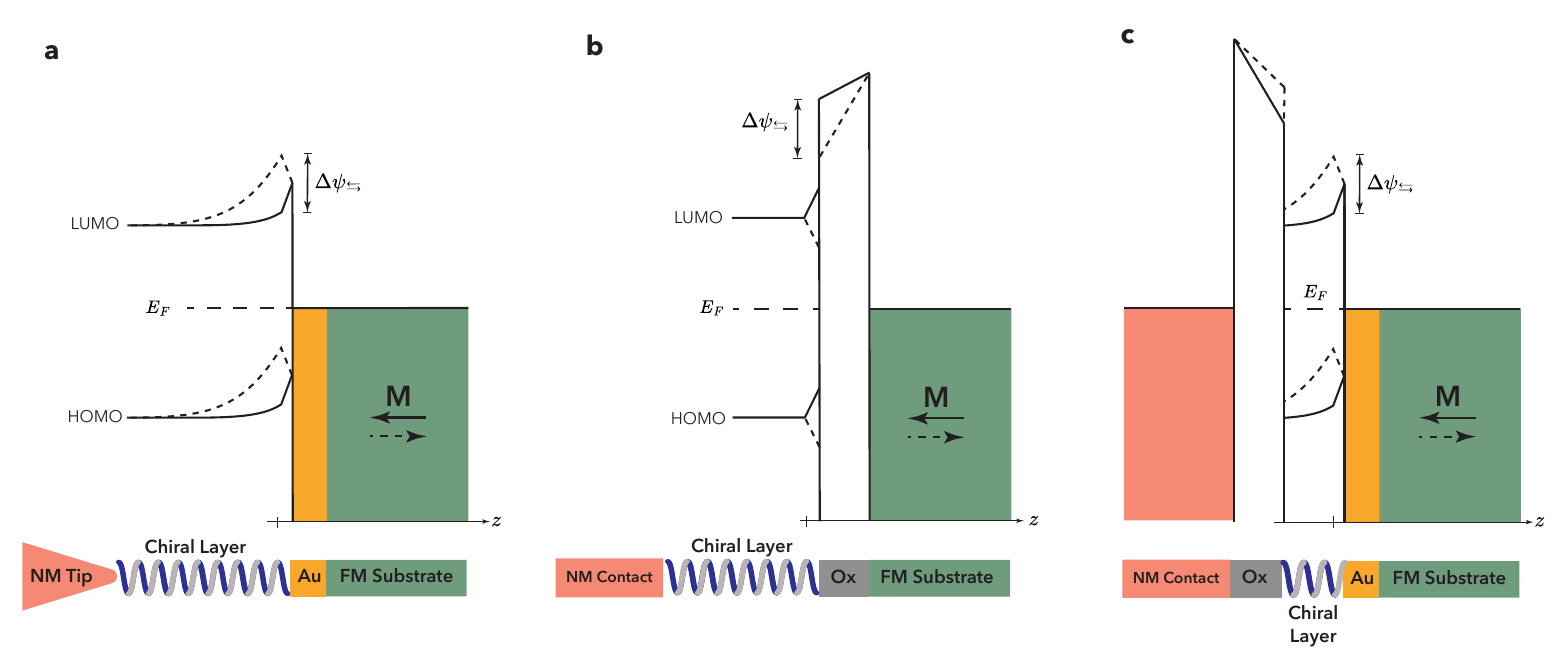}
\caption{\label{fig:tunnel_barier_modulation} Effect of the potential profile modification with a spacer between the ferromagnetic substrate and the HOMO/LUMO band of the chiral system. (a) In most experiments that use a ferromagnet substrate, there is a gold (Au) layer between the ferromagnet and the chiral system. However, this is not significantly different from the case where there is direct contact between the ferromagnet and the chiral system. (b) The potential profile of an oxide (Ox) tunnel barrier between the chiral system and the ferromagnet is also modified by the magnetization direction. Here, we ignore possible band bending of the HOMO/LUMO bands of the chiral system and show the potential profile only close to the ferromagnetic substrate, ignoring the normal metallic tip/substrate. (c) A proposed sample configuration with the chiral layer contacted by a gold coated ferromagnet on one side and by a tunnel barrier on the other side. In this geometry spin-dependent transport cannot generate an MR but the contact potential modification can.}
\end{figure*}

\begin{equation}
	\nabla ^2 \psi(z) = \frac{\partial^2 \psi(z)}{\partial z^2} = -\frac{\rho(z)}{\epsilon_s},
	\label{eq:PE eq}
\end{equation}

where $\psi(z)$ is the potential as function of the distance $z$ and $\rho(z)$ is the local electric charge density. With the boundary conditions $\psi(0) = \Delta\psi_\leftrightarrows$ and $\psi(\lambda) = 0$, the solution to Eq.~(\ref{eq:PE eq}) yields,

\begin{equation}
	\psi(z) = \frac{e N_d}{2 \epsilon_s}\left[ \lambda^{2} - ( \lambda - z)^2 \right],
	\label{eq:W_fm}
\end{equation}

here $N_D$ is the doping concentration in the chiral system and $\lambda$ is the screening length, given by,

\begin{eqnarray}
\lambda &= \sqrt{\frac{2 \epsilon_s \Delta\psi_{\leftrightarrows}}{e N_D}}
\;,
\qquad \Delta\psi_{\leftrightarrows} > k_B T,
\label{eq:T_R_Left_Right}
\\
\lambda &= \sqrt{\frac{ 2 \epsilon_s k_B T}{e^2 N_D}}
\;,
\qquad \Delta\psi_{\leftrightarrows} < k_B T.
\label{eq:T_R_Right_Left}
\end{eqnarray}

In FIG.~\ref{fig:workfunc_modulation_graph}(a) we plot $\psi(x)$ for the case where $\Delta \psi_{\leftrightarrows} > k_B T$ with a contact potential modification of \SI{50}{\milli \volt} at $N_D =$ \SI{e15}{\per\centi\meter\cubed} to illustrate the length scales on which the electrostatic potential is modified. The length $\lambda$, is dependent on $N_D$, as seen in the inserted graph in FIG.~\ref{fig:workfunc_modulation_graph}(a). From the graph of $\psi(z)$, it can be seen that the length scale of the potential modification is $\approx$ \SI{150}{\nano\meter}, making it comparable to if not greater than the molecular thin film thickness, which is typically less than \SI{20}{\nano\meter}. This illustrates that the potential landscape of the entire molecular thin film can be modified by the reversal of the magnetization direction, which will strongly alter the charge transport, also in linear response.

When the ferromagnet-chiral molecule junction is under bias, the contact Fermi energy gets shifted by $eV$. Three different mechanisms can carry the transport; see FIG.~\ref{fig:workfunc_modulation_graph}(b). In the first mechanism, (1) thermionic emission, the carriers need to have a comparatively high energy to overcome the barrier. The second transport mechanism is (2) quantum mechanical tunneling through the barrier, which is extremely sensitive to the potential landscape of the barrier. The third mechanism is (3) transport in the region within the screening length due to modification of the occupation of the states in the bands.

 Since the magnetization direction modifies the electrostatic potential landscape as well as the barrier height, all three transport mechanisms will be strongly influenced by the magnetization direction. Furthermore, we emphasize that both the transport via the HOMO and LUMO bands is modified by the magnetization direction. When the barrier height is increased, transport via the LUMO is reduced, but the opposite is true for transport via the HOMO. Therefore, the sign of the MR is expected to change when the doping or the carrier type (electrons or holes) is changed in the chiral system.

Two different experimental configurations are used for the local probe measurements. In one, the tip is ferromagnetic, and the substrate is non-magnetic, as is illustrated in FIG.~\ref{fig:workfunc_modulation}(c). In the other configuration, illustrated in FIG.~\ref{fig:tunnel_barier_modulation}(a), the substrate is ferromagnetic, and in almost all experiments covered by a thin (2 to 10 nm) gold (Au) layer, and the tip is non-ferromagnetic. Relevant here is experimental work by \citeauthor{Ghosh2020} \cite{Ghosh2020}, where the effect of an Au spacer layer between the ferromagnet and the chiral molecules was investigated. It was found that the shift in contact potential induced by reversing the magnetization direction is around \SI{40}{\milli\volt} for a \SI{10}{\nano\meter} thick Au layer, a comparatively large distance. Furthermore, it was found that reduction of the Au thickness increases the contact potential shift to more than \SI{100}{\milli\volt}. Therefore these two seemingly different configurations yield similar results (but note our final remarks in the conclusions and outlook section). We estimate that the potential landscape is modified on a length scale comparable to that of the thickness of chiral film and therefore this electrostatic effect can dominate the electronic (but not the spin) transport.

With a tunnel barrier between the chiral molecular system and the ferromagnet, the modification of the contact potential adjusts the potential landscape as is illustrated in FIG.~\ref{fig:tunnel_barier_modulation}(b), where we neglect possible bending of the HOMO/LUMO bands. The modified contact potential also alters the potential profile and height of the tunnel barrier. Interestingly, the magnetization direction that corresponds to an increased barrier height for a Schottky type junction (dashed lines in FIG.~\ref{fig:tunnel_barier_modulation}(a)) corresponds to a decrease in the average tunnel barrier height (dashed line in FIG.~\ref{fig:tunnel_barier_modulation}(b)). Some experimental results, where local probe and tunnel junction experiments are performed, give an opposite sign of the MR for the same chiral system and seem to support this \cite{Mishra2020b,Zhang2023}. However, it is of vital importance to make the comparison between Schottky barrier type junctions and tunnel junctions with the same definition of up and down magnetization, which is not always clear in the literature.

\section{Conclusions and Outlook}
\label{sec:con_out}
We have provided an alternative explanation for the origin of the linear response MR of chiral systems coupled to a ferromagnet. Experimental results show a shift in the contact potential of chiral molecules in proximity to a ferromagnet when reversing magnetization direction. We argue that the transport barrier between the chiral molecules and the ferromagnet is modified by the magnetization direction in equilibrium without any bias current. The length scale on which the potential is modified is of similar magnitude, if not greater than the thickness of the measured chiral thin film. Therefore, the electronic transport is strongly modified by the magnetization direction, resulting in an (large) MR in the linear response regime.

With this simplified model we propose an alternative explanation for the MR of chiral system coupled to a ferromagnet that does not require spin-dependent transport and explains the failure of the Onsager reciprocity relations. However, we emphasize that in principle nonlinear/strongly out of equilibrium directional spin-dependent transport in chiral systems can lead to an MR \cite{Yang2020a} (see Appendix \ref{sec:MR_CISS_non-linear}) purely based on spin transport.
  
 A number of questions remain to be addressed. The first is the origin of the contact potential modification by the magnetization direction, as of yet it is not understood what physical mechanism causes this, in particular in equilibrium. A (yet unknown) magnetic exchange mechanism could be responsible for the chirality dependent contact potential modification in the case of direct contact between the chiral system and the ferromagnet. It could also be applied to the gold coated ferromagnets. However the effect was shown to extend over \SI{10}{\nano\meter} of gold, a comparatively large distance for which the mechanism is unclear. It is also unclear how an exchange mechanism could be applied to the tunnel barrier that separates the chiral system and the ferromagnet.

The second remaining question concerns the bias independent nature of the MR at biases even up to several Volts. For a larger bias both the Schottky barrier type and the tunnel barrier potential profiles will be adjusted by the bias induced electric field. How the MR stays relatively independent of the applied bias is a question that remains to be answered. 

The third question is if and how our explanation can be applied to other types of experiments (beyond two terminal configurations) where chiral systems are coupled to a ferromagnet. A recent experiment has investigated the MR measured from the conduction in the plane of a ferromagnetic/chiral molecule bilayer \cite{Kondou2022}. A chirality-dependent asymmetry in the MR of the ferromagnet is found that is independent of the applied current, showing that coupling between (molecular) chirality and ferromagnetism can induce an effect on the measured MR that is not current dependent. This could also be relevant for measurements of adhesion forces between chiral systems and ferromagnets (with different magnetization directions) \cite{Ziv2019,Kapon2023a} or in field-effect transistor devices with a chiral channel \cite{Volpi2023}. This shows that the electrostatic interactions between chiral systems and ferromagnets can be a widely applicable phenomena with important fundamental physics to be explored as well as possible application in nanotechnology.

To finish, we identify five main approaches to further confirm the proposed mechanism for the MR. A systematic study should be made where the shift in contact potential is compared to the Schottky barrier height extracted temperature dependent electronic transport measurements. A second approach is to change the carrier type in the chiral system by means of doping or a gate electrode because we expect that the MR changes sign when the transport in HOMO changes to the LUMO. As a third approach we propose a yet to be explored geometry where the chiral material is directly contacted by a ferromagnet on one side and by an oxide barrier and nonmagnetic electrode on the other, as can be seen in FIG.~\ref{fig:tunnel_barier_modulation}(c). In this geometry any possible modulation of spin transport cannot generate a MR but the electrostatic potential modification can generate an MR which allows the two origins to be separated. The fourth approach is to investigate the transport behaviour of chiral systems in non local device geometries where the charge and spin transport can be fully decoupled. For instance in non local devices \cite{Yang2019} where ferromagnetic electrodes inject spin and electrodes made of a chiral system to detect the spin signal. Last but not least, a comparison of FIG.~\ref{fig:workfunc_modulation}(c) and FIG.~\ref{fig:tunnel_barier_modulation}(a) shows that the electrostatic modification is, for the same chirality and magnetization direction of the system, opposite for a magnetic electrode on top or below the chiral system. This implies that an experiment with a magnetic tip or a magnetic substrate would show MR with opposite signs.

\section{acknowledgments}

This work is supported by the Zernike Institute for Advanced Materials (ZIAM), and the Spinoza prize awarded to Professor B.J. van Wees by the Nederlandse Organisatie voor Wetenschappelijk Onderzoek (NWO). The authors like to acknowledge useful discussions during the Gordon Research Conference 'Electron Spin Interactions with Chiral Molecules and Materials' in August 2023. In particular with P.M. Theiler. We also thank M.H. Diniz Guimaraes and C.H. van der Wal for useful discussion and comments on the manuscript.

\appendix

\section{Linear response regime}
\label{sec:App_Linear_Resp}

In linear response, an applied bias $eV$ only causes a gradient/difference in the electrochemical potential that drives electron transport. Therefore the charge and spin transport parameters of the sample such as the conductivity can be fully determined by the equilibrium properties of the sample. This implies that the associated current induced electric field does not matter for the transport and that other current induced effects, such as the spin transfer torque, are not relevant in linear response \cite{Hedegard2023}. A conservative energy range for the linear response regime is $eV \approx k_bT$, where $k_b$ is the Boltzmann constant and $T$ is the temperature \cite{datta_1995}. However, $k_bT$ is a lower bound estimate; the linear response regime can extend to energies far above $k_bT$ e.g. if inelastic relaxation processes take place in the sample.

Owing to microscopic reversibility, the linear response follows several fundamental symmetry properties, most notably the Onsager reciprocity relations \cite{Buttiker1986}. As a consequence of the Onsager reciprocity relations, a system with a magnetization ($M$) in an external magnetic field ($B$) requires the linear response resistance to satisfy:

\begin{equation}
	R_{ij,nm}(B,M) = R_{nm,ij}(-B,-M),
	\label{eq:R4T_Ons_Butt}
\end{equation}

where the subscripts $ij$ correspond to the current contacts and $nm$ to the voltage contacts of the sample. In a two terminal connection this reduces to,

\begin{equation}
	R_{2T}(B,M) = R_{2T}(-B,-M),
	\label{eq:R2T_Ons_Butt}
\end{equation}

such that the resistance is unchanged when the magnetization and external magnetic field are reversed, meaning no MR in linear response. Note that this is under the assumption that only the magnetization is reversed and not possible associate electric fields. A disadvantage of a two-terminal measurement is that in addition to the sample resistance also the resistance of the contacts will be measured, significantly increasing the risk of spurious effects being detected. However, in the linear response regime, the resistance of the contacts also satisfies Eq.~(\ref{eq:R4T_Ons_Butt}) \& (\ref{eq:R2T_Ons_Butt}).

The Onsager reciprocity relations are very well established and have been tested in many experiments \cite{Leturcq2006}, including systems with strong spin-orbit interaction \cite{Lee2018a}, organic single-molecule junctions \cite{Mitra2022} and when a Büttiker (dephasing) probe is included \cite{Huisman2021,Varela2023}.

We are not aware of any experimental results nor theory that deviates from Eq.~(\ref{eq:R4T_Ons_Butt}) \& Eq.~(\ref{eq:R2T_Ons_Butt}) except the MR measured in chiral systems coupled to a ferromagnet (see main text).

\begin{figure*}
\centering
\includegraphics[width=0.99\textwidth]{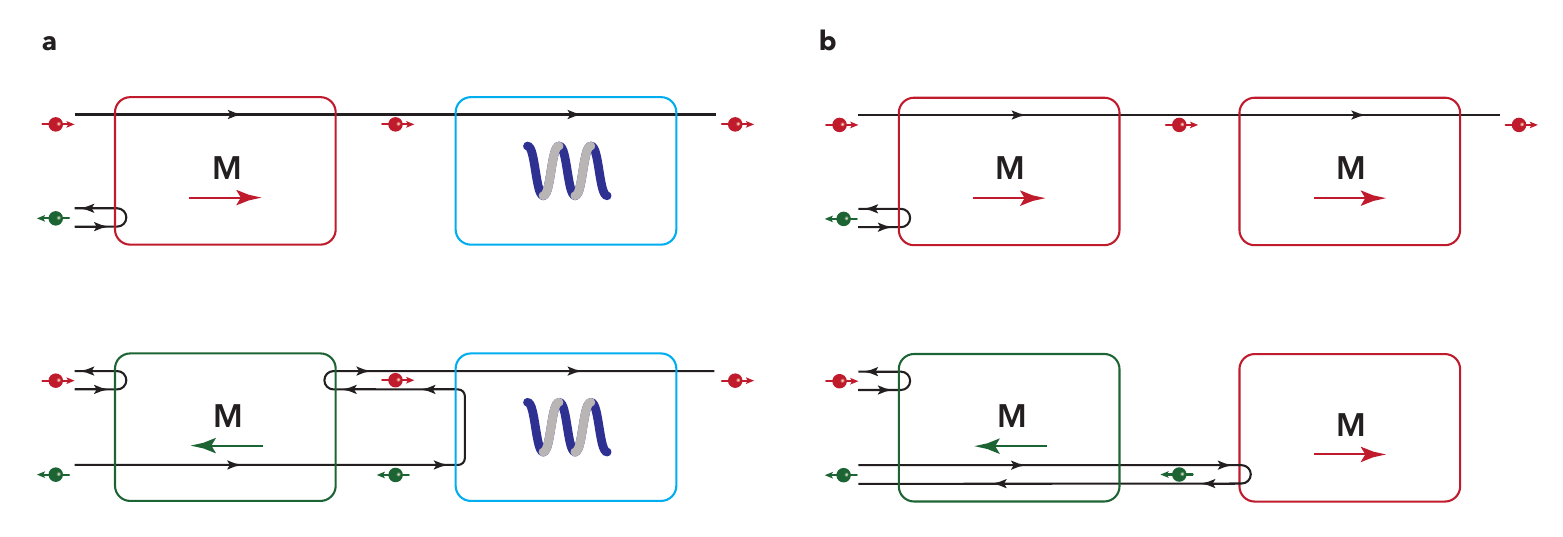}
\caption{\label{fig:App} (a) Spin-dependent electron transmission model for electrons propagating from left-to-right though a chiral system coupled to a ferromagnet. The reversal of the magnetization does not affect the total charge current and hence Eq.~(\ref{eq:R2T_Ons_Butt}) does hold. Note there are spin currents on both sides of the system which depend on the chirality and the magnetization direction but that the charge transport is unaffected by the magnetization direction, see Ref \cite{Yang2020a}. (b) This is fundamentally different from two connected ferromagnets and the magnetization of one of them being reversed.}
\end{figure*}

\section{Spin-dependent transmission for linear response magnetoresistance}

\label{sec:App_Linear_Resp_CISS}

It is key to distinguish the MR generated by the entire circuit of chiral system plus ferromagnet from the possible mechanisms which cause the spin-dependent electron transport in chiral systems, because they are fundamentally different questions. Unfortunately, the two concepts are sometimes mixed up.

The absence of MR in the linear response for a chiral system coupled to a ferromagnet was illustrated with a general spin-dependent transmission model developed by \citeauthor{Yang2019} \cite{Yang2019,Yang2020a,Yang2021}. This model is based on spin-dependent transmission and reflection coefficients as illustrated in FIG.~\ref{fig:workfunc_modulation}(a)\&(b). The ``favoured'' spin is transmitted and the ``unfavoured'' spin gets reflected with spin-flip, \cite{Yang2019} which is required by the absence of spin currents in equilibrium. Owing to the spin-dependent transport, spin currents are generated in the chiral system when it is biased, due to the spin-dependent transport in the chiral system. However, this does not result in a MR when the chiral system is connected to a ferromagnet. Transport through a ferromagnet is also spin-dependent. However, the crucial difference is that the magnetization direction sets the direction of the favoured spin, see FIG.~\ref{fig:workfunc_modulation}(b). 

FIG.~\ref{fig:App}(a) shows spin-dependent transport in a chiral system coupled to a ferromagnet, including spin transmission, reflection and spin-flip, for both magnetization directions. When the ferromagnet transmits electrons that are favoured by the chiral system, they are transmitted by the chiral system. With the magnetization reversed, the ferromagnet transmits electrons unfavored by the chiral system. However, owing to the spin-flip reflection process of the chiral system and the spin-conserving reflection of the ferromagnet, the total charge current is unchanged by the magnetization reversal, giving a strict zero MR in the linear response regime. Although the illustrated electron transmission and reflection in FIG.~\ref{fig:App}(a) assumes an ideal spin-dependent transport for both the ferromagnet and the chiral system, the same conclusion holds for a any ferromagnet and chiral system.

Fundamentally different is the spin valve consisting of two ferromagnets, where the total number of transmitted electrons is affected when the magnetization of only one of the ferromagnets is reversed, as can be seen in FIG.~\ref{fig:App}(b).  When the magnetizations of both ferromagnets are reversed the total resistance is unaffected, and the two terminal Onsager relation of Eq.~(\ref{eq:R2T_Ons_Butt}) is satisfied.

We note that electron transport in chiral crystals has been studied in devices where the current flows along a crystallographic axis \cite{Inui2020,Shishido2021,Shiota2021} and perpendicular to the current direction a contact with strong spin-orbit interaction measures a voltage proportional to the spin signal from the inverse spin Hall effect in the contact. For these devices, that do not have a magnetic electrode, the reciprocal measurement where voltage and current contacts are interchanged were observed to satisfy Eq.~(\ref{eq:R4T_Ons_Butt}) \cite{Inui2020,Shishido2021,Shiota2021}, the four terminal Onsager relations.

However, despite the low conductivity of many chiral systems hindering the studies of the transport at low bias, there are now several experiments that do show a clear linear response regime \cite{Torres-Cavanillas2020,Clever2022,AlBustami2022a,Bian2023,Malatong2023,Metzger2023}. Furthermore, they show a difference in the linear response resistance when the magnetization is reversed, giving a non-zero MR, contradicting the Onsager reciprocity relations. We note that, in a number of these results, the bias dependence of the MR shows anomalous behaviour around zero bias voltage \cite{Kulkarni2020,Bian2023,AlBustami2022a,Labella2023,Malatong2023,Zhang2023}. We attribute this to possible offset effects in the applied bias voltage and/or the measured current which can lead to incorrect experimental determination of the MR around zero bias.

Based on the discussion above and in section \ref{sec:tech_and_results}, we conclude that the experimental results of charge transport in chiral systems connected to a ferromagnet are inconsistent with the spin transport interpretation that is commonly used. This emphasises that an alternative explanation for the MR in the linear response is required, as we give in this paper.

\section{Linear response compared to non-linear response regime}
\label{sec:MR_CISS_non-linear}

In the linear response regime, the bias current does not significantly modify the properties of the sample compared to equilibrium of the system, and Onsager reciprocity relations are strict. It is important to note that the gradient in electrochemical potential drives the bias current and that the electric field is not relevant for the transport behaviour. This changes when the applied bias modifies the properties of the system under study and we enter the non-linear regime where the Onsager reciprocity relations are not strict. For a chiral system coupled to a ferromagnet, it was shown \cite{Yang2020a, Huisman2021} that, due to a combination of energy-dependent transport and energy relaxation, a MR can be generalized in the non-linear regime when the magnetization is reversed. Furthermore, the MR is strongly bias dependent, has to be strictly zero at zero bias and change sign when bias is reversed, meaning that the MR is an odd function in the bias.

This has similarity to the well-established electrical magnetochiral anisotropy (EMCA) \cite{Rikken2001,Rikken2023} where the resistance of a chiral system has an additional chirality-dependent ($D/L$) term $\Delta R$ that is proportional to the applied current $I$ and the magnetic field $B$,

\begin{equation}
	\Delta R = R_o^{D/L} I \cdot B
\end{equation}

, making the MR an odd function of $I$ and $B$.

However, this is entirely inconsistent with the experimental observations of transport in chiral systems coupled to a ferromagnet because they show a MR, which is, in many reports, independent of the applied bias, down to zero bias, making it an even function of the applied bias. 

\bibliography{Workfunc_modulation}

\end{document}